\documentclass[PRD,twocolumn,showpacs,superscriptaddress,preprintnumbers,nofootinbib,amsmath,amssymb]{revtex4}

\usepackage{graphicx}
\usepackage{dcolumn}
\usepackage{bm}
\usepackage{color}

\def\fnl{{f_{\rm{nl}}}}
\def\taunl{{\tau_{\rm{nl}}}}
\def\fnle{{\widehat \fnl}}
\def\taunle{{\widehat \taunl}}

\def\fnlet{{\widehat \taunl}}

\def\hnlett{{\taunle}}

\def\VEV#1{\left\langle #1 \right\rangle}

\newcommand{\beq}{\begin{equation}}
\newcommand{\eeq}{\end{equation}}
\newcommand{\beqa}{\begin{eqnarray}}
\newcommand{\eeqa}{\end{eqnarray}}

\newcommand{\Npix}{N_{\mathrm{pix}}}

\newcommand{\lmax}{{l_{\mathrm{max}}}}

\begin{document}

\title{The probability distribution for non-Gaussianity estimators constructed from the CMB trispectrum}

\author{Tristan L.~Smith}
\affiliation{Berkeley Center for Cosmological Physics, Physics
     Department, University of California, Berkeley, CA 94720} 
\affiliation{IPMU, University of Tokyo, 5-1-5 Kashiwanoha, Kashiwa, Chiba 277-8583, Japan}
\author{Marc Kamionkowski}
\affiliation{California Institute of Technology, Mail Code 350-17,
Pasadena, CA 91125}
\affiliation{Department of Physics and Astronomy, Johns Hopkins University, Baltimore, MD 21218, USA}

\date{\today}

\begin{abstract}
Considerable recent attention has focussed on the prospects to
use the cosmic microwave background (CMB) trispectrum to probe
the physics of the early universe.  Here we evaluate the probability
distribution function (PDF) for the standard estimator $\taunle$ for the
amplitude $\taunl$ of the CMB trispectrum both for the null-hypothesis (i.e., for Gaussian maps with $\taunl=0$) and for maps
with a non-vanishing trispectrum ($\taunl\neq0$). 
We find these PDFs to be {\it highly} non-Gaussian in both cases.
We also evaluate the variance with which the trispectrum
amplitude can be measured,  $\VEV{\Delta \widehat{\tau}_{\rm nl}^2}$, 
as a function of its underlying value, $\taunl$.   We find a strong
dependence of this variance on $\taunl$.  We
also find that the variance does not, given the highly
non-Gaussian nature of the PDF, effectively characterize the
distribution.  Detailed knowledge of these PDFs will therefore
be imperative in order to properly interpret the implications of
any given trispectrum measurement.  For example, if a CMB experiment with a 
maximum multipole of $\lmax = 1500$ (such as the Planck satellite) 
measures $\taunle = 0$ then at the 95\% confidence our calculations show that 
 we can conclude $\taunl \leq 1005$; assuming a Gaussian PDF but with the correct $\taunl$-dependent variance we would 
 incorrectly conclude $\taunl \leq 4225$; further neglecting the $\taunl$-dependence in the variance 
we would incorrectly conclude $\taunl \leq 361$.
\end{abstract}


\maketitle

\section{Introduction}

Current observations of the cosmic microwave background (CMB) and 
large-scale structure (LSS) provide a powerful probe of the physics of the 
early universe.  As an example, the near scale-invariance of the 
primordial power-spectrum along with an upper limit to the inflationary gravitational-wave background
can be used to rule out a few of the simplest models of inflation \cite{Finelli:2009bs}.  
A measurement of the statistics of the primordial perturbations 
can provide an even more discriminating 
test of models of the early universe: \emph{all} canonical single-field slow-roll inflation models predict that the 
perturbations are observationally indistinguishable from Gaussian \cite{inflation,localmodel}.
Therefore any observed deviation from Gaussianity will rule out all canonical 
single-field slow-roll
inflation models. 
However, non-canonical single-field models \cite{nonconon}, multi-field models \cite{larger}, curvaton 
models \cite{curvaton}, and models with sharp features \cite{Wang:2000} or
wiggles \cite{Hannestad:2009yx} may produce larger departures from 
Gausianity.  Measurement of the level of primordial non-Gaussianity has thus become one of
the primary goals of CMB and
LSS research.  

The majority of efforts to measure primordial non-Gaussianity from the CMB have relied on an estimator
constructed from the bispectrum, the three-point correlation
function in harmonic space \cite{Luo:1993xx}.  However, most models that predict a 
significant level of non-Gaussianity also
predict a non-zero connected trispectrum (the non-Gaussian part of 
the harmonic-space four-point function)
\cite{Kunz:2001ym,Hu:2001fa,Okamoto:2002ik,DeTroia:2003tq,Kogo:2006kh,Regan:2010cn},
and some efforts have been mounted to detect a primordial non-Gaussian signature 
from the trispectrum \cite{Kunz:2001ym,DeTroia:2003tq,Smidt:2010ra}.  In this way a constraint 
on the trispectrum amplitude provides unique information 
on a broad range of early-universe processes such as multi-field inflation
models \cite{NfinlationTRI}, the curvaton senario \cite{isocurvTRI}, inflation models with 
non-standard kinetic terms \cite{nonStandardKinTRI}, and
the influence of primordial cosmic strings \cite{string_trispectrum}.

The level of non-Gaussianity is often quantified using the `local-model' through
the non-Gaussianity parameter $\fnl$ defined by
\cite{Luo:1993xx},
\begin{equation}
     \Phi(\vec x) = \phi(\vec x) + \fnl \left[\phi(\vec x)^2-\VEV{\phi^2}\right],
\label{fnldefinition}
\end{equation}
where $\Phi(\vec x)$ is the curvature potential and $\phi(\vec x)$ a
Gaussian random field.  Current limits from the
CMB/LSS constrain the value to be $|\fnl|\lesssim 80$ at 95\% confidence level (c.l.)
\cite{limits,halos}.  The Planck satellite \cite{:2011ah} is
expected to achieve a sensitivity of $\fnl\sim5$.  

Constraints on the amplitude of the non-Gaussian local-model CMB bispectrum and trispectrum have very 
broad implications. Although 
various physical processes predict a range of values for $\fnl$, 
it can be shown that \emph{all} single-field models of inflation predict \cite{Creminelli:2004yq}
\begin{equation}
\fnl = \frac{1}{4}(n_s-1), 
\end{equation}
where $n_s$ is the slope of the primordial power spectrum.  
Current constraints to $n_s$ \cite{Keisler:2011aw} imply that 
all single-field models predict $\fnl \simeq -0.008$.  Therefore, 
if the Planck satellite constrains $\fnl$ to be non-zero,
we will be able to make the profound statement that all single-field models  
are disfavored by the data.  
A measurement of the amplitude of the local-model trispectrum\footnote{The local-model trispectrum can be defined by using 
 Eq.~(\ref{fnldefinition}) with the identification $\taunl =
\fnl^2$.}, $\taunl$, 
may lead to an additional test of the basic cosmological model.  Recently 
Ref.~\cite{Smith:2011if} has shown that $\taunl$ respects the inequality
\begin{equation}
\taunl \geq \frac{1}{2} (\fnl)^2, 
\label{eq:taunlconsis}
\end{equation}
independent of the underlying physics.  A constraint on both $\fnl$ and $\taunl$ using the CMB may appear to 
violate Eq.~(\ref{eq:taunlconsis}) at the expense of actually violating translation invariance \cite{Smith:2011if}.  Therefore, a constraint on both 
$\fnl$ and $\taunl$ provides a very broad test of some of the fundamental assumptions in our standard cosmological model. 
Given the wide-ranging impact of constraints on $\fnl$ and $\taunl$, it is of great importance to 
report the significance of any constraint accurately. 

 To date, most studies which use CMB observations 
to place constraints on $\fnl$ and $\taunl$ have used estimators that are constructed to have the minimum 
variance under the null hypothesis
\cite{Okamoto:2002ik,Babich:2004yc,Hu:2001fa}.  In order to use
these estimators to place meaningful 
constraints on $\fnl$ and $\taunl$ we 
must know the full shape of their probability 
distribution functions (PDFs) as a function of $\fnl$, $\taunl$,
and the maximum multipole $\lmax$ of the observation.
  Thus,
for example, we often evaluate or forecast the standard error
$\sigma$ with which a given measurement will recover the
true value of $\fnl$ and $\taunl$ and then simply assume that the error is
Gaussian.  If so, then with $\sigma_{\taunl}=100$, for example, a
measurement of $\taunle=300$ would represent a $3\sigma$ departure
from $\taunl=0$, and a measurement $\taunle=0$ would represent a
$3\sigma$ departure fom $\taunl=300$.  However, if the PDF depends
on the true value $\taunl$, and if that distribution is
non-Gaussian, then it may be that a measurement $\taunle=300$ could
be easily consistent with a true value $\taunl=0$, while a
measurement $\taunle=0$ could be inconsistent with $\taunl=300$ with
a confidence greater than ``$3\sigma$.''   We will see
below that something like this occurs with measurements
of $\taunl$.

A calculation of these PDFs is particularly important for measurements of
non-Gaussianity (as opposed, for example, for the CMB power
spectrum), because the estimator is a sum over products of three (in the 
case of $\fnle$) or four (in the case of $\taunle$)
temperature measurements.  This is unlike the power spectrum which sums
over squares of temperature measurements.  Suppose the
temperature is measured in $\Npix$ pixels.  There are then $\sim
\Npix^2$ terms in the $\fnl$ estimator
and $\sim
\Npix^3$ terms in the $\taunl$ estimator (after restrictions
imposed by statistical isotropy).  While these terms may have
zero covariance, they are not statistically independent; there
is no way to construct $\Npix^2$ or $\Npix^3$ statistically independent
quantities from $\Npix$ measurements. The conditions required
for the validity of the central-limit theorem are therefore not
met, and the estimators will not necessarily approach a Gaussian in the 
$\Npix \gg 1$ limit.

Although previous 
studies have calculated the PDF in the case of $\fnl$ \cite{Creminelli:2006gc, Smith:2011rm}, 
the only property of the $\taunl$ estimator that has been explored in 
the literature is how the
variance for the null-case scales with the maximum observed 
multipole, $\lmax$ \cite{Okamoto:2002ik,Kogo:2006kh,Smidt:2010ra}.  
In order to address how well CMB observations can estimate $\taunl$ 
we calculate the PDF $P[\taunle; \taunl,\lmax]$---the
probability that a given measurement with resolution $\lmax$
will return a value $\taunle$ given that the underlying theory
has a value $\taunl$---using numerous Monte Carlo realizations
of an ideal (no-noise) flat-sky map in the Sachs-Wolfe 
approximation.   In order to both generate the maps and apply the estimator to them we use 
a fast-Fourier transform (FFT) algorithm described in Appendix A of Ref.~\cite{Smith:2011rm}.  
Lessons learned about
$P[\taunle; \taunl,\lmax]$ in this ideal case help to interpret and understand
current/forthcoming results and assess the validity of
full-experiment simulations.  

Our simulations show that $P[\taunle; \taunl,\lmax]$ is highly non-Gaussian for all 
values of $\taunl$, including the null case.  Additionally, our simulations 
allow us to derive, for the first time, how the variance of this distribution depends on the underlying value of $\taunl$.  
Neglecting this dependence, Ref.~\cite{Kogo:2006kh} concludes
that the signal-to-noise\footnote{The signal-to-noise  is
defined to be $S/N \equiv \taunl/\sigma_{\taunl}$.  In the case
of Gaussian noise with a 
variance independent of $\taunl$ this is a measure of the fractional error in a constraint to $\taunl$.  However, 
in the case of $\taunle$, the noise is neither Gaussian nor
independent of $\taunl$ so that the quantity
$\taunl/\sigma_{\taunl}$ is only an approximation to the
significance of a constraint on $\taunl$.} of
this estimator appears to scale as $\taunl^2$, becoming more sensitive to non-Gaussianity for large $\taunl$
than an estimator using the CMB bispectrum.
We show that the dependence of the variance on the underlying value of $\taunl$ is significant, finding
that the sensitivity of this 
estimator to local-model non-Gaussianity is always weaker than
that of the bispectrum  estimator, and it approaches a constant
for $\taunl \gtrsim 10^9/l^2_{\rm max}$. 

Knowledge of both the variance and shape of $P[\taunle; \taunl,\lmax]$ is necessary to 
assign proper confidence levels (c.l.) to 
constraints.  For example, if the Planck satellite ($\lmax \simeq 1500$) measures 
 $\taunle = 0$ then at the 95\% c.l.~our calculations show that 
 we can conclude $\taunl \leq 1005$.  If we assumed a Gaussian PDF but with the correct $\taunl$-dependent variance we would 
 incorrectly conclude $\taunl \leq 4225$.  If we also neglected to 
 include the $\taunl$-dependent variance 
we would incorrectly conclude $\taunl \leq 361$. 

This paper is organized as follows.  In
Sec.~\ref{tri_formalism}, we discuss how to 
construct the minimum-variance estimator $\taunle$ using the CMB 
trispectrum under the null-hypothesis.   In
Sec.~\ref{sec:localPDF} we apply this estimator to the 
local-model for non-Gaussianity.  In Sec.~\ref{sec:nullPDF}
we present our results for the PDF in the null ($\taunl = 0$) case.  Sec.~\ref{sec:nonnullPDF} presents our 
results for the non-null case and gives a fitting formula for $P[\taunle; \taunl,\lmax]$.  In Sec.~\ref{sec:discussion} we summarize our results. 

\section{Non-Gaussianity estimators constructed from the CMB trispectrum}
\label{tri_formalism}
\subsection{Formalism}

We assume a flat sky to avoid the complications (e.g., spherical
harmonics, Clebsch-Gordan coefficients, Wigner 3$j$ and 6$j$
symbols, etc.) associated with a spherical sky, and we further
assume the Sachs-Wolfe limit.  We denote the fractional
temperature perturbation at position $\vec\theta$ on a flat sky
by $T(\vec\theta)$ and refer to it hereafter simply as the
temperature.

The field $T(\vec\theta)$ has a power spectrum
$C_l$ given by
\begin{equation}
     \VEV{T_{\vec l_1} T_{\vec l_2}} = \Omega \delta_{\vec
     l_1+\vec l_2,0} C_l,
\label{eqn:powerspectrum}
\end{equation}
where $\Omega=4\pi f_{\mathrm{sky}}$ is the survey area (in
steradians), 
\begin{equation}
     T_{\vec l} = \int\, d^2\vec \theta\, e^{-i\vec l\cdot
     \vec \theta} T(\vec\theta) \simeq
     \frac{\Omega}{N_{\mathrm{pix}}} \sum_{\vec\theta} e^{-i\vec l\cdot
     \vec \theta} T(\vec\theta),
\end{equation}
is the Fourier transform of $T(\vec\theta)$, and $\delta_{\vec
l_1+\vec l_2,0}$ is a Kronecker delta that sets $\vec l_1 =
-\vec l_2$.  In the limit of small departures from Gaussianity, $C_l$ is also the power spectrum
for $T(\vec\theta)$, which for a scale-invariant primordial
power spectrum with amplitude $A$ ($A \simeq 10^{-10}$) is given by
\begin{equation}
     C_l = \frac{2 \pi A}{l^2}.
\end{equation}

The trispectrum is defined by \cite{Hu:2001fa,Okamoto:2002ik}
\begin{equation}
     \VEV{T_{\vec l_1} T_{\vec l_2} T_{\vec l_3}
      T_{\vec l_4}} = \taunl \Omega
     \delta_{\vec l_1 +\vec l_2 +\vec l_3+ \vec l_4,0} 
     {\cal T}(\vec l_1,\vec l_2,\vec l_3,\vec l_4),
\end{equation}
and for the local-model,
\begin{eqnarray}
          {\cal T}(\vec l_1,\vec l_2,\vec l_3,\vec l_4) &=& 
          P_{l_3l_4}^{l_1 l_2}(|\vec l_1+\vec l_2|) . \nonumber \\
          &+ & 
          P_{l_2l_4}^{l_1 l_3}(|\vec l_1+\vec l_3|) +
          P_{l_2l_3}^{l_1 l_4}(|\vec l_1+\vec l_4|) , \nonumber\\
\label{eqn:localtri}
\end{eqnarray}
where
\begin{eqnarray}
               P_{l_3l_4}^{l_1 l_2}(|\vec l_1+\vec l_2|) &\equiv&
               4 C_{|\vec l_1+\vec l_2|} \left[ C_{l_1} C_{l_3} + C_{l_1}
               C_{l_4} \right. \nonumber \\
               & & \left. + C_{l_2} C_{l_3} + C_{l_2}
               C_{l_4}\right].
\label{eqn:Pdefn}
\end{eqnarray}
Due to statistical isotropy, the trispectrum is nonvanishing only for $\vec l_1 +\vec
l_2 +\vec l_3+ \vec l_4=0$, that is, only for quadrilaterals in
Fourier space.

\subsection{The minimum-variance trispectrum estimator}
\label{sec:Triestimator}

Each distinct quadrilateral $\vec
l_1 + \vec l_2+\vec l_3 +\vec l_4=0$ gives an estimator for the
trispectrum with some variance.  Adding the individual
estimators with inverse-variance weighting gives the
minimum-variance estimator,
\begin{eqnarray}
     \fnlet &=& \sigma_{T,0}^{2} \sum_{\vec l_1+ \vec l_2+ \vec l_3+ \vec l_4=0} \frac{
     T_{\vec l_1} T_{\vec l_2}
     T_{\vec l_3} T_{\vec l_4} }{4!\Omega^3 C_{l_1}C_{l_2}C_{l_3}C_{l_4}}\label{eqn:triestimator} \\
     &\times&{\cal T}(\vec l_1,\vec l_2,\vec
     l_3,\vec l_4)- \sigma_{T,0}^{2}\VEV{\mathcal{T}}_{G},\nonumber
\end{eqnarray}
where we have subtracted off the unconnected (Gaussian) part of the 
trispectrum,  $\VEV{\mathcal{T}}_{G}$, and the inverse variance is given by
\begin{equation}
     \sigma_{T,0}^{-2} = \sum_{\vec l_1+ \vec l_2+ \vec l_3+ \vec l_4=0} \frac{ \left[{\cal T}(\vec
     l_1,\vec l_2,\vec l_3,\vec l_4) \right]^2}{4! \Omega^2
     C_{l_1} C_{l_2} C_{l_3} C_{l_4}},
\label{eqn:trivariance}
\end{equation}
where $2 \leq |\vec l_i| \leq \lmax $.

\section{The PDF of $\taunle$ for the local-model}
\label{sec:localPDF}

We now restrict our attention to the local family of
non-Gaussian models [see Eq.~(\ref{fnldefinition})] in which the
temperature $T(\vec{\theta})$ has a non-Gaussian component,
\begin{equation}
     T(\vec{\theta}) = t(\vec{\theta}) + 3 \sqrt{\taunl} \left\{ [t(\vec{\theta})]^2
     -\VEV{[t(\vec{\theta})]^2} \right\},
\label{eqn:localmodel}
\end{equation}
where we have chosen the normalization $\tau_{\rm nl}$ to correspond 
to amplitude of the non-Gaussian part of 
the Newtonian potential four-point function\footnote{This 
differs by a factor of $5/6$ the usual definition which is given in terms of the Bardeen 
potential \cite{Boubekeur:2005fj}.}.
 To zeroth order in $\taunl$, the power spectrum and
correlation function for $T(\vec \theta)$ are the same as those for
$t(\vec \theta)$.  Note that $T(\vec \theta)$ is, strictly speaking, the
temperature fluctuation, so $\VEV{T(\vec \theta)}=0=T_{\vec
l=0}$. 

The temperature Fourier coefficients can be written $T_{\vec
l}=t_{\vec l}+\sqrt{\taunl} \delta t^2_{\vec l}$ with
\begin{eqnarray}
     \delta t^2_{\vec l} &\equiv & \frac{3}{\Omega} \sum_{\vec
     l_1 + \vec l_2 = \vec l } t_{\vec l_1} t_{\vec l_2}.
     \label{eq:nonGauss}
\end{eqnarray}
Formally, the sum goes from $1 \leq |\vec l_1| < \infty$, but
for a finite-resolution map, the sum is truncated at some $l_{\rm
max}$ such that the number of Fourier modes equals the number of
data points.

Applying the estimator in Eq.~(\ref{eqn:triestimator}) to the local-model 
trispectrum [Eq.~(\ref{eqn:localtri})], we obtain
\begin{eqnarray}
   \taunle = 2  \sum_{1\leq|\vec L|\leq2 l_{\rm max}} 
  C_L \Bigg| \sum_{\vec l_1+\vec l_2+ \vec L = 0}
  	\frac{T_{\vec l_1} T_{\vec l_2}}{\Omega^2C_{l_1}}\Bigg|^2
	-\sigma_{T,0}^{2}\VEV{\mathcal{T} }_{G},
	\label{eq:trilocal}
\end{eqnarray}
where we can now write
\begin{equation}
\VEV{\mathcal{T}}_{G}= 2 \sigma_{T,0}^2 \sum_{|\vec L| > 1}C_L 
\sum_{\vec l_1 + \vec l_2 + \vec L=0} \left(\frac{C_{l_1}}{C_{l_2}} + 1\right).
\end{equation}

\subsection{The PDF of $\hnlett$ under the null-hypothesis, $\taunl = 0$}
\label{sec:nullPDF}

Under the null hypothesis ($\taunl = 0$) we apply $\hnlett$ to a
purely Gaussian CMB temperature map.
\begin{figure}[htbp]
\resizebox{!}{8cm}{\includegraphics{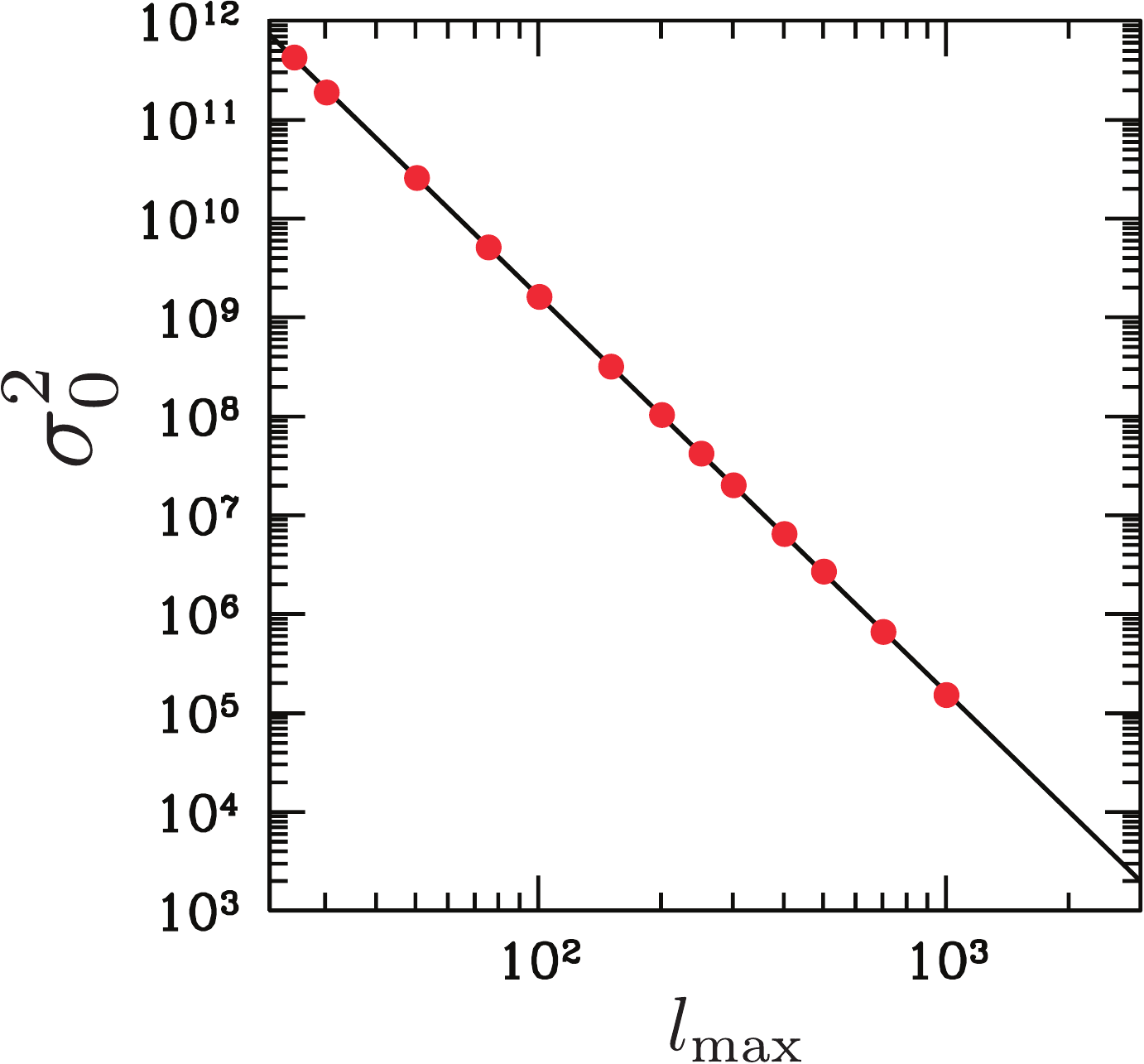}}
\caption{The scaling of the variance of $\hnlett$ under the null
hypothesis $\taunl = 0$.  Each red 
point shows the result of a Monte Carlo simulation for 1000 realizations.  The black line is 
a fit to those simulations given by Eq.~(\ref{eq:sigma0fit}).}
\label{fig:nullVarScaling}
\end{figure}
As shown in Fig.~\ref{fig:nullVarScaling}, our Monte 
Carlo simulations find that the variance $\sigma_0$ of this
estimator under the null hypothesis as a function of the maximum
multipole $l_{\rm max}$ included in the analysis is well-fit 
by a power-law
\begin{equation}
\sigma^2_0(\lmax) = \frac{1.74 \times 10^{-2}}{A^2 \lmax^4}.
\label{eq:sigma0fit}
\end{equation}
This scaling compares well with the results of previous work \cite{Okamoto:2002ik,Kogo:2006kh}.

\begin{figure}[htbp]
\resizebox{!}{12cm}{\includegraphics{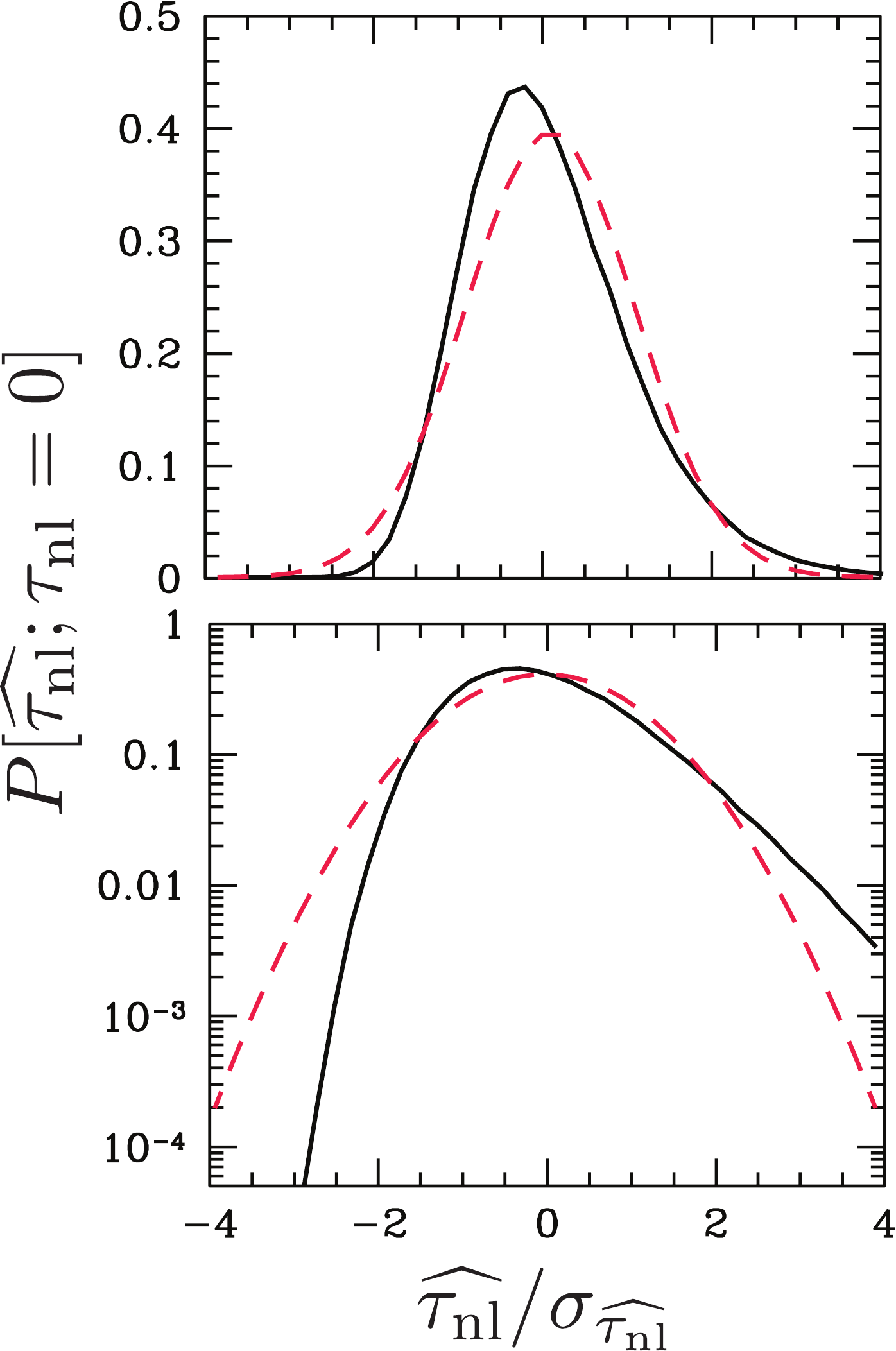}}
\caption{The black-solid curve shows the PDF of $\hnlett$ under 
the null-hypothesis, $\taunl = 0$ for $l_{\rm max} = 50$ and calculated 
with $10^6$ realizations.  The red-dashed curve shows a Gaussian 
PDF with the same variance.  The upper panel shows the PDF 
on a linear scale, the low panel on a logarithmic scale.  When scaled by its variance, the PDF is 
identical for $l_{\rm max} = 100$ showing that it takes on a 
universal form in the $l_{\rm max} \gg 1$ limit.  We give a fitting formula 
for this PDF in Eq.~(\ref{eq:fitting}).}
\label{fig:nullPDF}
\end{figure}

The simulations also allow us to calculate the full shape of the PDF of this estimator under the 
null hypothesis.  Since 
the number $\Npix$ of measurements of the CMB temperature map is
much less than the number ($\sim \Npix^3$)
of terms in this estimator, the standard 
central-limit theorem does not apply so that the PDF
$P[\hnlett; \taunl = 0,\lmax]$, is not necessarily Gaussian in 
the $\Npix \gg 1$ limit.  The simulations demonstrate that 
the PDF is, in fact, highly non-Gaussian as shown in Fig.~\ref{fig:nullPDF}.  
The asymmetry of the PDF can be explained by referring to the expression 
for the estimator in Eq.~(\ref{eq:trilocal}).  There it can be seen that the estimator 
is bounded from below but unbounded from above. 
By calculating the PDF, $P[\hnlett; \taunl = 0,\lmax]$, for several 
values of $l_{\rm max}$ we find that when scaled by its variance it
 takes on the universal shape shown by the black curve in Fig.~\ref{fig:nullPDF}.  
The PDF is well fit by the formula,
\begin{eqnarray}
   &&P[ \hnlett; \taunl = 0, \lmax] = \frac{1}{\sigma N} \label{eq:fitting}\\
 &&    \begin{cases}
      e^{-\frac{1}{2}\big|\frac{\hnlett/\sigma-x_p}{ \sigma_p}\big|^n}, & \hnlett/\sigma \leq x_p,\\
      e^{-\frac{c}{\sigma_p^2}\left(\sqrt{(\hnlett/\sigma-x_p) ^2 + c^2}-c\right)}, & \hnlett/\sigma > x_p,
\end{cases}\nonumber
\end{eqnarray}
where $\sigma$ is the variance of the estimator, given in Eq.~(\ref{eq:sigma0fit}), 
the normalization $N$ is given by
\begin{equation}
N \equiv 2 \sigma_p \Gamma 
\left(\frac{n+1}{n}\right)+c\ e^{c^2/\sigma_p^2} K_1\left(\frac{c^2}{\sigma_p^2}\right), 
\end{equation}
where $\Gamma$ is the Euler Gamma function, and $K_1$ is the
modified Bessel function of the first kind.  We find that the PDF is best fit by the parameters $n=3$,
$x_p = -0.13$, $\sigma_p = 0.64$, and $c= 0.488$.

\subsection{The PDF of $\hnlett$  with  $\taunl \neq 0$}
\label{sec:nonnullPDF}

For $\taunl \neq 0$ the non-Gaussianity in the CMB map imparts further non-Gaussianity to the shape of 
$P[\hnlett; \taunl,\lmax]$.  In addition to this, the variance has a strong dependence on $\taunl$ so that 
when $\taunl$ and $\lmax$ are large enough the ratio $\taunl/\sigma_{\taunl}$, which approximates the $S/N$ of the estimator, approaches a constant value. 

 In order to investigate how the variance of $\hnlett$ depends on $\taunl$ it is useful to expand it in 
 powers of $\taunl$.  Given that $T_l$ is linear in $\taunl$ and that $\hnlett$ is quartic in $T_l$ the 
 expansion includes terms up to $\taunl^2$:
\begin{equation}
\hnlett = \mathcal{T}_0 + \sqrt{\taunl}  \mathcal{T}_1 + \taunl  
\mathcal{T}_2 + \taunl^{3/2}  \mathcal{T}_3 + \taunl^2  \mathcal{T}_4,
\label{eq:triexpand}
\end{equation}
where each $ \mathcal{T}_i \sim \Sigma t_l^{i+4}$.  We give explicit expressions for the $\mathcal{T}_i$ in 
Appendix A. 

Since only cross-correlations which include even products of $t$
are non-zero, the  variance of $\hnlett$ is given by
\begin{eqnarray}
\left \langle \left[\Delta \hnlett\right]^2\right \rangle &=& \sigma_0^2 + \taunl (\sigma_1^2 + \sigma_{0,2}^2) 
\\ &+&
\taunl^2 (\sigma_2^2 + \sigma_{0,4}^2+\sigma^2_{1,3})
\nonumber \\&+& \taunl^3 (\sigma_3^2+\sigma_{2,4}^2) + \taunl^4 \sigma_4^2,\nonumber
\label{eq:TriVar}
\end{eqnarray}
where, for example, $\sigma_{0,2}^2$ denotes the covariance 
between $\mathcal{T}_0$ and $\mathcal{T}_2$ and $\Delta \taunle \equiv \taunle-\VEV{\taunle}$. 

\begin{figure}[htbp]
\resizebox{!}{8cm}{\includegraphics{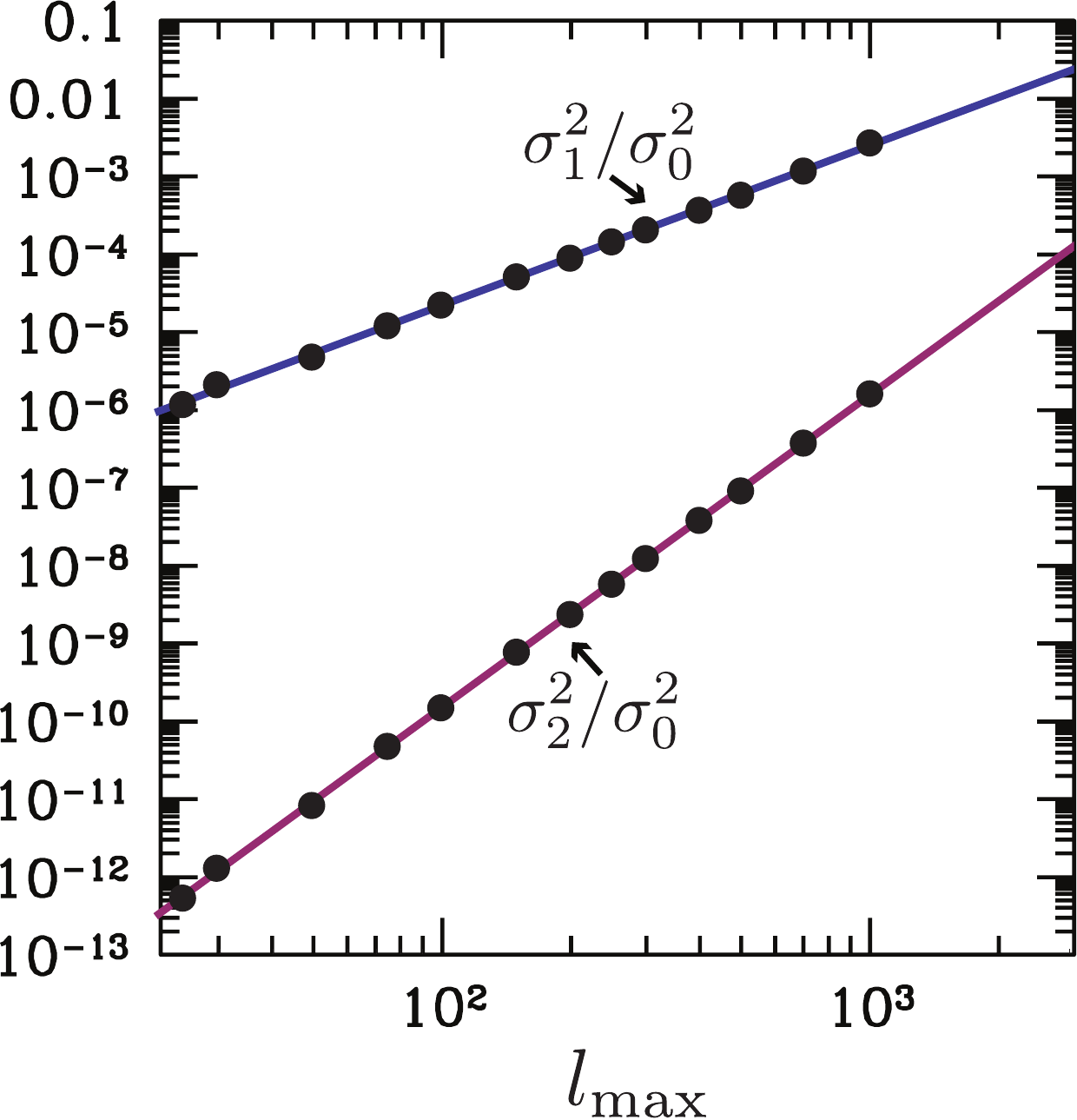}}
\caption{The variances $\sigma_1^2$ and $\sigma_2^2$ as a fraction of the 
zeroth-order variance $\sigma_0^2$.  The points are the results of the Monte Carlo 
simulations and the curves show the power-law fits given in Eqs.~(\ref{eq:sigma1}) and (\ref{eq:sigma2}). }
\label{fig:var_scaling}
\end{figure}

Calculating the terms that appear in Eq.~(\ref{eq:TriVar}) using 
 Monte Carlo simulations we find that for $\taunl < 10^4$ and $l_{\rm max} <10^4$ only $\sigma_1^2$
and $\sigma_2^2$ significantly contribute.  The variance is then well approximated by
\begin{equation}
\left \langle \left[\Delta \hnlett\right]^2\right \rangle \approx
\sigma_0^2 + \taunl \sigma_1^2 +
\taunl^2 \sigma_2^2,
\label{eq:TriVar2}
\end{equation}
where $\sigma_0^2$ is given in Eq.~(\ref{eq:sigma0fit}) and 
\begin{eqnarray}
\sigma_1^2 &=& \frac{0.028}{A \lmax^2}, \label{eq:sigma1}\\
\sigma_2^2 &=& 0.23. \label{eq:sigma2}
\end{eqnarray}
The results of our Monte Carlo simulations are shown in Fig.~\ref{fig:var_scaling}.

\begin{figure}[htbp]
\resizebox{!}{8cm}{\includegraphics{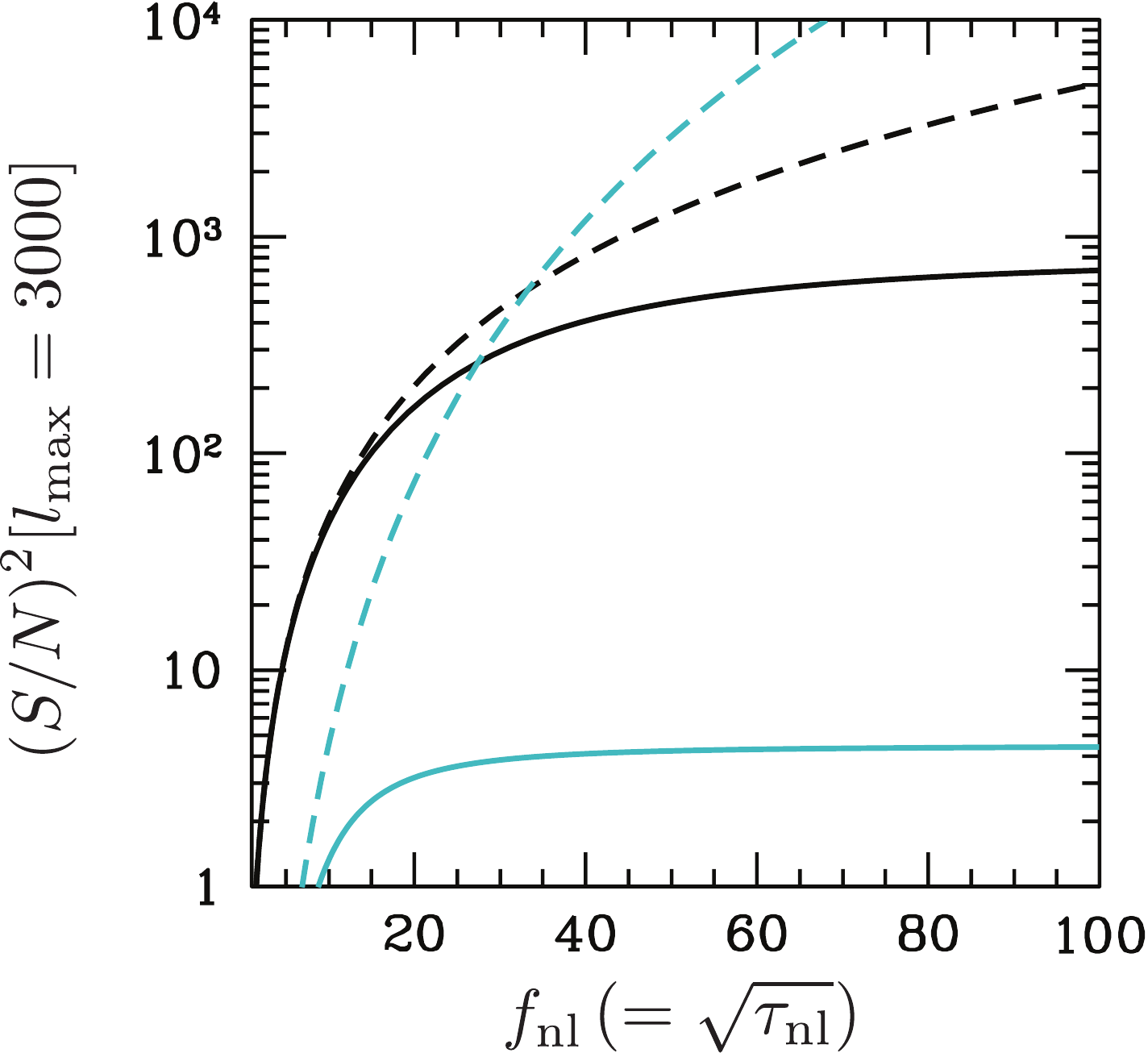}}
\caption{The ratio $\fnl^2/\sigma^2_{\fnl}$ using the CMB bispectrum (black, from Ref.~\cite{Smith:2011rm}) and 
$\taunl^2/\sigma^2_{\taunl}$ using the CMB trispectrum [blue, Eq.~(\ref{eq:TriVar2})] as a function of $\taunl = \fnl^2$.  
These ratios can be interpreted as an estimate for the $S/N$ 
for a constraint to local-model non-Gaussianity in the CMB. The dashed curves show the scaling of the $S/N$ without 
taking into account 
the dependence of the variance on $\fnl$ and $\taunl$; the solid curves show 
the correct $S/N$ scaling.  As in Ref.~\cite{Kogo:2006kh}, from the dashed curves we would (incorrectly) 
conclude that the trispectrum estimator 
is more sensitive to a non-Gaussian signal for $\sqrt{\taunl} \gtrsim 40$.}
\label{fig:SN_scaling}
\end{figure}

The scaling given in Eq.~(\ref{eq:TriVar}) shows that for a large enough $\lmax$ the 
variance of the estimator scales 
as $\taunl^2$ so that the ratio $\taunle/\sigma_{\taunle}$ becomes constant for $\taunl \gtrsim 0.1/(A \lmax^2)$.  A similar scaling is 
observed with the minimum-variance null-hypothesis 
estimator using the CMB bispectrum \cite{Creminelli:2006gc, Smith:2011rm}.  
Neglecting the dependence of the variance on $\taunl$, previous work \cite{Kogo:2006kh} claimed 
that for large enough $\taunl$ the minimum-variance null-hypothesis estimator using the 
CMB trispectrum would be more sensitive to a local-model non-Gaussian signal than an 
estimator using the CMB bispectrum with the relationship $\taunl = \fnl^2$.  Given the 
dependence of the variance on 
$\taunl$ our calculations demonstrate that this is not the case, 
as shown by the solid curves in Fig.~\ref{fig:SN_scaling}. 

When $\hnlett$ is applied to a map with $\taunl \neq 0$ then 
the non-Gaussianity in the map 
imparts additional non-Gaussianity to $P[\hnlett; \taunl,\lmax]$.  We are 
interested in calculating the shape of the PDF for an experiment such as 
Planck which has $\lmax \simeq 1500$.  Although, in principle, it is possible 
calculate the PDF for large $\lmax$, it is computationally demanding especially 
given the large number of realizations we must generate in order to explore the 
tails of the distribution.  The 
computation can be simplified since  
we find the PDF is `self-similar' in the sense that its shape depends on the 
ratios $\sigma_0^2/\sigma_1^2$ and $\sigma_0^2/\sigma_2^2$.  
Given Eqs.~(\ref{eq:sigma0fit}), (\ref{eq:sigma1}), and (\ref{eq:sigma2}) 
this implies that the shape depends on the combination $\taunl l^2_{\rm max}$.  

\begin{figure}[htbp]
\resizebox{!}{8cm}{\includegraphics{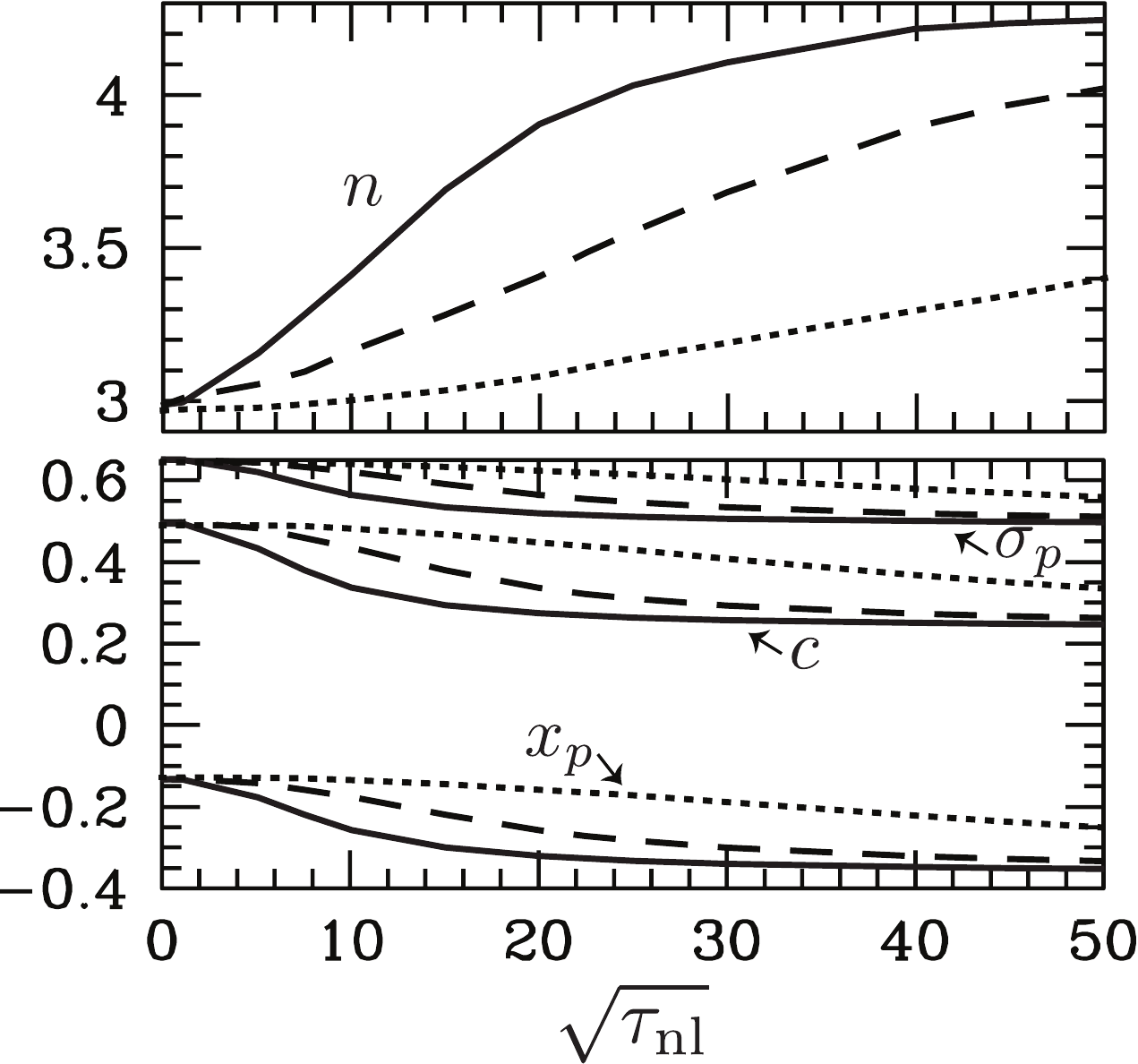}}
\caption{The scaling of the parameters for the fitting formula in 
Eq.~(\ref{eq:fitting}) with $\sqrt{\taunl}$ for  $\lmax = 600$ (dotted curves) $\lmax = 1500$ (dashed curves) 
and $\lmax = 3000$ (solid curves) .  For $\taunl \neq 0$ the 
non-Gaussianity of the map imparts additional non-Gaussianity to the PDF.  
As $\taunl$ increases the asymmetry of the PDF increases 
with the power-law index of the PDF for $\hnlett < \taunl$ going from $n=3$ to $n=4$ in the 
case of $\lmax = 1500$, and $n=4.24$ in the case of $\lmax =3000$.}
\label{fig:param_scaling}
\end{figure}

Using this fact it is straightforward to calculate the PDF for a moderate 
value of $\lmax$ (we used $\lmax = 50$) and then scale the PDF to a 
larger value for various choices of $\taunl$.  We find 
that $P[\hnlett; \taunl,\lmax]$ is well fit by the formula used to fit $P[\hnlett; \taunl=0,\lmax]$, 
given by Eq.~(\ref{eq:fitting}), with parameters $n$, $\sigma_p$, 
$x_p$, and $c$ which now depend on $\fnl$ and $\sigma$ is the variance of the 
estimator given by Eq.~(\ref{eq:TriVar2}).  In Fig.~\ref{fig:param_scaling} we show how 
these parameters depend on $\taunl$ for $\lmax = 600$ (dotted), $\lmax = 1500$ (dashed) and $\lmax = 3000$ (solid).  We find that 
as $\taunl$ increases the asymmetry of the PDF increases 
with the power-law index of the PDF for $\hnlett < \taunl$ growing from $n=3$ to $n=4$ for $\lmax = 1500$ and $n=4.24$ for $\lmax = 3000$.

Our knowledge of the full shape of $P[\hnlett; \taunl,\lmax]$ now allows us to properly 
assign confidence levels (c.l.).  If a given CMB observation with $\lmax$ yields a value of $(\taunle)_{\rm obs}$ we 
assign the 95\% c.l.~by finding the value of $(\taunl)_{\pm \rm c.l.}$ which satisfies the integral equation
\begin{equation}
0.95 = \bigg| \int_{(\taunle)_{\rm obs}}^{(\taunle)_{\pm}}
P[\hnlett; (\taunl)_{\pm \rm c.l.},\lmax] d\hnlett \bigg|,
\end{equation}
where $(\taunle)_{\pm}$ is the solution to the equation 
\begin{equation}
P[(\taunle)_{\pm}; (\taunl)_{\pm \rm c.l.},\lmax] = P[(\taunle)_{\rm obs}; (\taunl)_{\pm \rm c.l.},\lmax].
\end{equation}

If, for example, an experiment with $\lmax = 3000$ measures 
 $\VEV{\hnlett} = 100$ then at the 95\% c.l., $\taunl = 100^{+210}_{-64}$.  If we assumed 
 a Gaussian PDF with the $\taunl$-dependent variance given 
by Eq.~(\ref{eq:TriVar2}) we would incorrectly conclude $\taunl= 100^{+3180}_{-94}$.  
Finally, if we also neglected to include the $\taunl$-dependent variance 
we would incorrectly conclude $\taunl = 100 \pm 92$.  

\section{Discussion}
\label{sec:discussion}

In this paper we have shown that the PDF for non-Gaussianity estimators using the 
CMB trispectrum
cannot be assumed to be Gaussian, since the number, $\sim\Npix^3$, of terms
used to construct these estimators greatly exceeds the number
$\Npix$ of measurements.  The 99.6\% confidence-level
interval cannot safely be assumed to be 3 times the 68.2\%
confidence-level interval.  We found that the PDF of standard
minimum-variance estimator $\hnlett$ using the CMB trispectrum constructed under
the null hypothesis is well-approximated by a distribution given
by Eqs.~(\ref{eq:fitting}) and (\ref{eq:TriVar2}).
This distribution is exponentially suppressed for values
$\hnlett < \taunl$ and enhanced for values $\hnlett > \taunl$,
relative to a Gaussian with the same variance.  We calculated
how the parameters of this fitting formula depend on $\taunl$
for $\lmax = 600$, $\lmax = 1500$ and $\lmax = 3000$ as shown in
Fig.~\ref{fig:param_scaling}.
We also find that the non-Gaussianity of $P[\hnlett; \taunl,\lmax]$ is greater for $\taunl \neq 0$.  

We have calculated, for the first time, how the variance of
$\hnlett$ depends on the underlying value of $\taunl$, as shown in Eq.~(\ref{eq:TriVar2}).  Previous 
work neglected this dependence leading to the incorrect conclusion that 
for large enough $\taunl$ and $\lmax$ a non-Gaussianity estimator constructed from the CMB trispectrum 
would have a larger $S/N$ than an estimator constructed from the CMB bispectrum \cite{Kogo:2006kh}.  
When the $\taunl$ dependence in the variance is included the $S/N$ of the estimator constructed from the 
CMB trispectrum becomes constant for $\taunl > 0.1/(A l^2_{\rm max})$.  As a result, the 
estimator constructed from the CMB bispectrum always 
produces a larger $S/N$, as shown by the solid curves in Fig.~\ref{fig:SN_scaling}.

These results have important consequences for future constraints to $\taunl$ measured 
from the CMB.  As discussed in the Introduction, future constraints to both $\fnl$ and 
$\taunl$ may imply wide-ranging conclusions about the physics of the early universe.  In 
particular, a non-zero measurement of $\fnl$ would rule out \emph{all} single-field inflation models and a constraint to $\taunl$ probes basic physical assumptions of the early universe, such as translation invariance, by
testing the consistency relation, $\taunl \geq \fnl^2/2$ around the surface of last scattering \cite{Creminelli:2004yq, Smith:2011if}. 
 If we suppose an experiment with $\lmax = 3000$ measures 
$\fnle = 20$ and  
$\taunle = 0$ then at the 95\% c.l.~our calculations show that 
we can conclude $\taunl \leq 200$, violating the consistency relation $\taunl \geq \fnl^2/2$ at the 95\% c.l.  If
we assumed a Gaussian PDF for $\taunle$ but with the correct
$\taunl$-dependent variance found in Eq.~(\ref{eq:TriVar2}) we
would  incorrectly find $\taunl \leq 1000$, leading to the false
conclusion that $\taunl \geq \fnl^2/2$ is consistent with the data.  If we also neglected to 
 include the $\taunl$-dependent variance 
we would incorrectly find $\taunl \leq 90$.

The non-Gaussian shape of the PDF of $\taunle$ is important even for current constraints.  The current published constraints on $\taunl$ from the Wilkinson Microwave Anisotropy Probe (WMAP) \cite{limits,Smidt:2010ra} ($\lmax=600$)
 are $\taunl/10^4 = 1.68 \pm 1.31$ at the 68\% c.l.~\cite{Smidt:2010ra}.  The error quoted in this constraint 
is estimated without taking into account the full shape of the PDF.  Although our results are not directly applicable 
to this case since they do not take into account several details of the WMAP analysis (such as a full CMB transfer function 
and the noise properties of the observations) they do allow us to discuss how using the correct PDF would qualitatively change 
the confidence levels.  For $\lmax = 600$ our calculations show that when assuming a Gaussian PDF along with a $\taunl$ independent variance 
the constraint is $\taunl/10^4 = 1 \pm 0.1$; the full PDF shows the actual constraint to be $\taunl/10^4 = 1 ^{+1.61\ + 3.3}_{-0.2\ -0.6}$.  
This indicates that a complete treatment of the confidence levels for the current constraint on $\taunl$ is both asymmetric and has a 
larger range than the constraint that is quoted in Ref.~\cite{Smidt:2010ra}.

The results presented here are made within the flat-sky, 
Sachs-Wolfe approximation.  As such our conclusions should be taken as an 
order-of-magnitude estimate of $P[\hnlett; \taunl,\lmax]$ calculated on the full 
sky and with the full transfer function (see Ref.~\cite{Creminelli:2006gc} for 
a further discussion).  However, we note that a comparison between 
the exact and approximate scaling of the $S/N$ with $\lmax$ shows the agreement 
to be better than an order of magnitude \cite{Babich:2004yc}. 

In this paper we have concentrated solely on the non-Gaussian local-model, defined in Eq.~(\ref{fnldefinition}). 
Although the quantitative results will differ for other models of non-Gaussianity, we expect that the 
qualitative conclusions will remain unchanged.  The non-Gaussian PDF for $\hnlett$ results from a breakdown 
of the central-limit theorem due to the large number of terms used in the estimator compared 
to the number of independent measurements.  Therefore, the work 
presented here shows that estimators for the CMB trispectrum
amplitude cannot be assumed to have a Gaussian PDF.  Rather, one
must carefully explore the full shape of the PDF before
assigning the significance of any particular
measurement.  Similar considerations may also need to be
considered, for example, in measurements of things like weak
gravitaitonal lensing, departures from statistical isotropy
\cite{Joshi:2011vc}, and the like, as the magntiudes of many of
these effects are determined in practice by the trispectrum.

\begin{acknowledgments}
TLS was supported by the 
Berkeley Center of Cosmological Physics and thanks the Institute for the Physics and Mathematics of the Universe (IPMU), University of Tokyo, for 
hospitality while this work was completed. 
MK was supported by DoE DE-FG03-92- ER40701 and NASA NNX12AE86G. 
\end{acknowledgments}

\begin{appendix}

\section{The expansion of $\hnlett$ in $\taunl$}
\label{sec:expansion}

Here we write down the explicit formulas for the expansion of $\taunle$, the
minimum-variance estimator using the CMB trispectrum constructed
under the null hypothesis, as 
written out schematically in Eq.~(\ref{eq:triexpand}).  The standard estimator applied to the 
non-Gaussian local-model, defined by Eq.~(\ref{fnldefinition}), can be written
\begin{eqnarray}
&&\hnlett +\sigma_{T,0}^{2}\VEV{\mathcal{T}}_{G}= \\ 
&& 2 \sigma_{T,0}^{2} \sum_{|\vec L|=-2 l_{\rm max},\ \vec L\neq0}^{2 l_{\rm max}} 
C_L \Bigg| \sum_{\vec l_1+\vec l_2+ \vec L = 0}
  	\frac{T_{\vec l_1} T_{\vec l_2}}{\Omega^2C_{l_1}}\Bigg|^2\nonumber.
\end{eqnarray}
Noting that $T_{\vec l} = t_{\vec l} + \sqrt{\taunl} \delta t^2_{\vec l}$ we have 
\begin{eqnarray}
\hnlett &=&  2 \sigma_{T,0}^{2} \sum_{|\vec L|=-2 l_{\rm max},\ \vec L\neq0}^{2 l_{\rm max}} C_L \bigg\{ |A_1|^2 + 
\sqrt{\taunl}  \left( A_1 A_2^* + A_2 A_1^2\right)\nonumber \\ &+& \taunl 
\left(|A_2|^2+ A_1 A_3^* + A_3 A_1^*\right) \nonumber\\
&+& \taunl^{3/2}\left(A_2 A_3^* + A_3 A_2^*\right) + \taunl^2 |A_3|^2 \bigg\}-\VEV{ \hnlett } \bigg |_{\taunl = 0},
\end{eqnarray}
where 
\begin{eqnarray}
A_1(\vec L) &\equiv& \sum_{\vec l_1 + \vec l_2 + \vec L = 0} \frac{t_{\vec l_1} t_{\vec l_2}}{C_{l_1}}, \\
A_2(\vec L) &\equiv& \sum_{\vec l_1 + \vec l_2 + \vec L = 0} 
\frac{t_{\vec l_1} \delta t^2_{\vec l_2}+t_{\vec l_2} \delta t^2_{\vec l_1}}{C_{l_1}}, \\
A_3(\vec L) &\equiv& \sum_{\vec l_1 + \vec l_2 + \vec L = 0} 
\frac{\delta t^2_{\vec l_1} \delta t^2_{\vec l_2}}{C_{l_1}}.
\end{eqnarray}

\end{appendix}

\end{document}